\documentclass[journal]{IEEEtran}

\usepackage{amssymb, graphicx, cite, amsmath, psfrag, minibox, mathrsfs, bigints, stfloats, color, upgreek, gensymb, epstopdf, textcomp, amsthm, amsfonts, euscript, calligra}

\usepackage[font=footnotesize]{subcaption}
\usepackage[font=footnotesize]{caption}

\usepackage{mathtools, amsfonts, multirow, array}
\usepackage{verbatim, mdwtab, relsize, cleveref, mdwmath}
\usepackage{algorithmicx}
\usepackage{algpseudocode}
\usepackage[ruled]{algorithm2e}

\usepackage{setspace}
\usepackage{float}

\usepackage{tikz}
\usetikzlibrary{shapes,arrows}
\usepackage{textcomp}

\def\pioh{\widehat{\pi}_0}
\def\pilh{\widehat{\pi}_1}
\def\P0{P^{(0)}}

\def\SUTx{SU$_{\text{tx}}$}
\def\SURx{SU$_{\text{rx}}$}
\def\PUTx{PU$_{\text{tx}}$}
\def\PURx{PU$_{\text{rx}}$}
\def\Ppk{{P}_{\text{pk}}}
\def\Ipk{{I}_{\text{pk}}}

\ifCLASSINFOpdf
\else
\fi

\begin{document}
\title{On Optimal Sensing and Capacity Trade-off in Cognitive Radio Systems with Directional Antennas}

\author{Hassan Yazdani,
        Azadeh~Vosoughi,~\IEEEmembership{Senior Member,~IEEE} \\
        University of Central Florida\\
        E-mail: {\tt \normalsize h.yazdani@knights.ucf.edu,  azadeh@ucf.edu} 
}

\markboth{}%
{Shell \MakeLowercase{\textit{et al.}}: Bare Demo of IEEEtran.cls for Journals}
%

\newcounter{MYeqcounter}
\multlinegap 0.0pt                     

\maketitle

 
\begin{abstract}
We consider a cognitive radio system, in which the secondary users (SUs) and primary users (PUs) coexist. The SUs  are equipped with steerable directional antennas. In our system, the secondary transmitter (\SUTx) first senses the spectrum (with errors) for a duration of $\tau$, and, then transmits data to the secondary receiver (\SURx) if spectrum is sensed idle. The sensing time as well as the orientation of \SUTx's antenna affect the accuracy of spectrum sensing and yield a trade-off between spectrum sensing and capacity of the secondary network. We formulate the ergodic capacity of secondary network which uses energy detection for spectrum sensing. We obtain optimal \SUTx ~transmit power, the optimal sensing time $\tau$ and the optimal directions of \SUTx ~transmit antenna and \SURx ~receive antenna  by maximizing the ergodic capacity, subject to peak transmit power and outage interference probability constraints.  Our simulation results show the effectiveness of these optimizations to increase  the ergodic capacity of the secondary network.
\end{abstract}
%
%

\IEEEpeerreviewmaketitle

\section{Introduction}\label{Se1}
%
The explosive rise in demand for high data rate wireless applications has turned the spectrum into a scarce resource. Cognitive radio (CR) technology is a promising solution which alleviates spectrum scarcity problem by allowing an unlicensed (secondary) user to access licensed bands in a such way that its imposed interference on the primary users (PUs) is limited \cite{Arslan}.  The focus of most literature is optimizing spectrum sensing and transmission strategies for opportunistic spectrum access of secondary users (SUs), when the SUs are equipped with omni-directional antennas \cite{Nalla, Liang, R003, Beaulieu, Kiskani1, Yousefvand}. In those works, spectrum sensing seeks spectrum holes in the time domain so that SUs exploit them for transmitting their data. Different from the bulk of the literature, in this paper we assume the SUs are equipped with steerable directional antennas which allow them to use spatial spectrum holes \cite{ICASSPpaper, R04, Golbon , Asilomar1} to increase spectrum utilization, specially in cognitive satellite networks \cite{CogSat2}. The directional antennas can identify and enable transmission and reception across spatial domain  and further enhance spectrum utilization, compared with omni-directional antennas. 
\par
\begin{figure}[!t]
\vspace{-4mm}
\centering
\hspace{-0mm}
\begin{subfigure}[b]{0.25\textwidth}                
 \vspace*{-10pt}
\setlength{\unitlength}{2.4mm} 
\centering
\scalebox{0.75}{
\begin{picture}(27,9)
\hspace{-4mm}
\small
\put(-2.55,-0.5){\line(0,1){5}}
\put(-2.5,.5){\framebox(32,4){}}
\put(-2.0,2.1){Sensing}
\put(2.5,-0.5){\line(0,1){5}}
\put(3.0,2.1){Data Transmission}
\put(13.5,-0.5){\line(0,1){5}}
\put(14.0,2.1){Sensing}
\put(18.5,-0.5){\line(0,1){5}}
\put(19.0,2.1){Data Transmission}
\put(29.53,-0.5){\line(0,1){5}}

\put(-2.55,-0.35){\vector(1,0){5}}
\put(-2.55,-0.35){\vector(-1,0){0}}
\put(-0.2,-1.5){$\tau$}

\put(2.5,-0.35){\vector(1,0){11}}
\put(2.5,-0.35){\vector(-1,0){0}}
\put(6.7,-1.5){$T - \tau$}

\put(13.5,-0.35){\vector(1,0){5}}
\put(13.5,-0.35){\vector(-1,0){0}}
\put(15.8,-1.5){$\tau$}

\put(18.5,-0.35){\vector(1,0){11}}
\put(18.5,-0.35){\vector(-1,0){0}}
\put(22.7,-1.5){$T - \tau$}
\end{picture}}
\vspace{0mm} 
\caption{Frame structure of secondary users.} 
\label{Frame}
\vspace{3mm} 	         
  \end{subfigure} \\
  
  \begin{subfigure}[b]{0.5\textwidth}
  \centering
		\includegraphics[width=44mm]{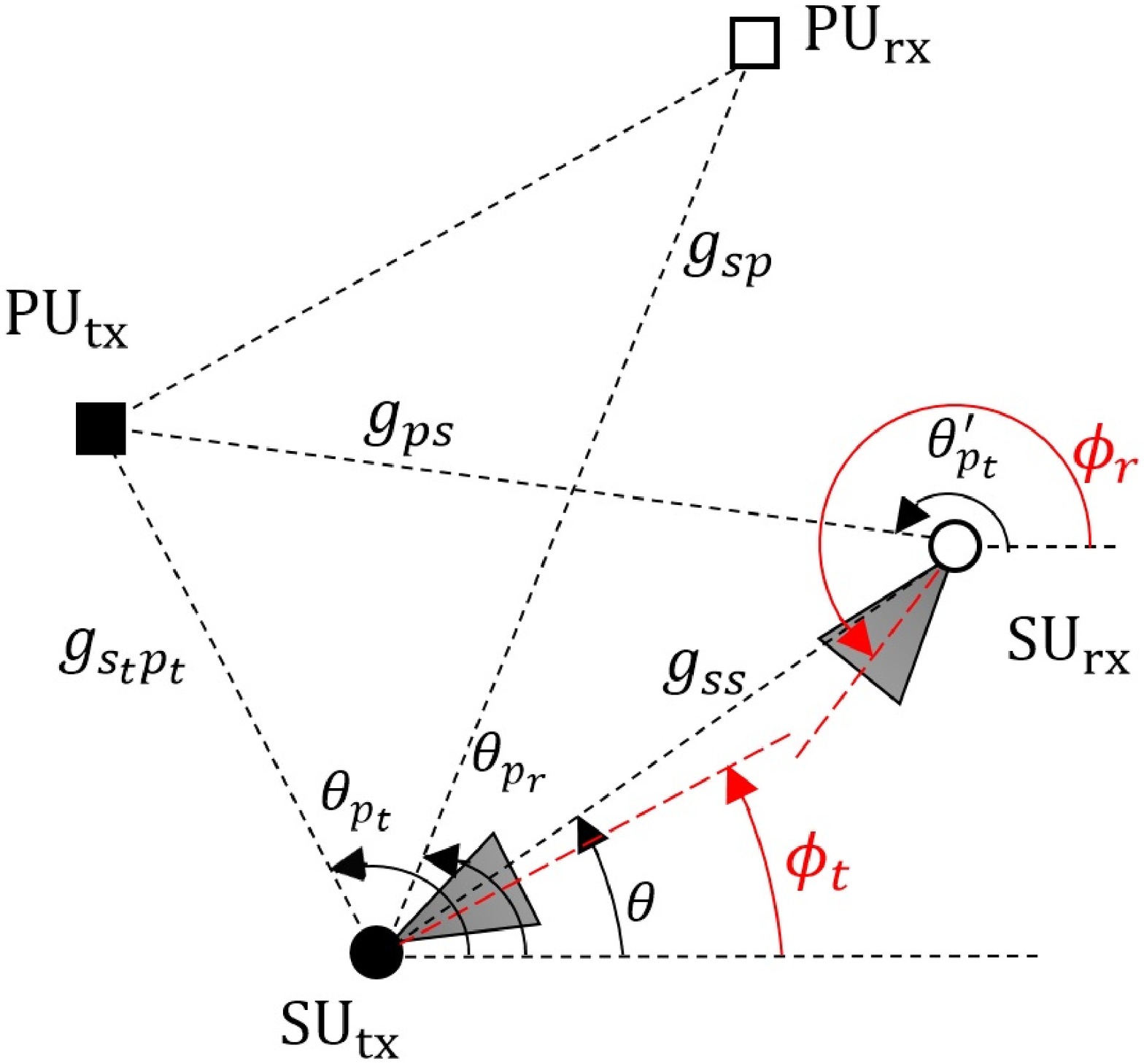}
		\caption{Our CR system with steerable directional antennas.} 
		\label{SystemModelFig}   
  \end{subfigure}
\caption{ Our system model.}
\vspace{-5mm}
\end{figure}

 In this paper, the SU transmitter (\SUTx) first senses the spectrum and  transmits data only when the spectrum is sensed idle. Since all spectrum sensing methods, including the energy detection method we use, are prone to sensing errors their false alarm and detection probabilities should be incorporated in the design and performance analysis \cite{Gursoy}. Suppose, the \SUTx ~employs a frame with duration $T$ seconds, depicted in Fig. \ref{Frame}, for spectrum sensing and data transmission. Each frame consists of a sensing time slot with duration $\tau$ seconds and the \SUTx ~uses this time  to decide whether spectrum is idle or busy. The remaining frame of duration $T-\tau$ seconds is used for data transmission if the spectrum is sensed idle. As $\tau$ increases, the false alarm probability decreases and detection probability increases. Thus, the result of  spectrum  sensing will be more accurate. On the other hand, the available time for data transmission decreases. Therefore, a trade-off exists between the sensing time and the capacity of our CR network.
\par  We assume that the \SUTx ~knows only the channel state information (CSI) of link between the \SUTx ~and the secondary receiver (\SURx), and the statistics of the other links.  Also, we assume that the \SUTx ~knows the geometry of CR network. The orientation of the \SUTx's antenna with respect to the direction of primary transmitter (\PUTx) affects the spectrum sensing accuracy. During spectrum sensing, to increase the detection probability and to receive the maximum power, the \SUTx's antenna should be pointed to the \PUTx's direction. On the other hand, the \SUTx's antenna should be pointed to the \SURx's direction to maximize the transmission capacity. Thus, in addition to sensing-capacity  trade-off in terms of sensing time $\tau$, there is  another sensing-capacity trade-off in terms of the \SUTx's antenna orientation. In this work, we establish the ergodic capacity of the channel between the \SUTx ~and the \SURx, when spectrum sensing is {\it imperfect} and find the optimal directions of the \SUTx ~and the \SURx ~antennas, the optimal \SUTx ~transmit power and the optimal sensing time $\tau$ such that the ergodic capacity is maximized, subject to two constraints, namely, peak transmit power and outage interference probability constraints.
{\color{red}
%
%
}
%
%
%
%
%
%
%
%
%
%
%
\section{System Model}\label{Se2}

\subsection{Network Geometry}
\par Our CR system model is shown in Fig. \ref{SystemModelFig}. The SUs are equipped with steerable directional antennas. The orientation of primary receiver (\PURx) with respect to \SUTx ~is denoted by $\theta_{p_r}$. Also, the orientation of \PUTx ~and \SURx ~with respect to \SUTx ~are denoted by $\theta_{p_t}$ and $\theta$, respectively and the orientation of \PUTx ~with respect to \SURx ~is denoted by $\theta'_{p_t}$.  The orientation of SU$_{\text{tx}}$ and SU$_{\text{rx}}$ antennas in their local coordination are denoted by $\phi_t$ and $\phi_r$, respectively (to be optimized). We assume $\theta_{p_t}$, $\theta_{p_r}$, $\theta$ and $\theta'_{p_t}$ are known or can be estimated \cite{MILCOM}. The  antenna gain is modeled as $A(\phi) = A_1 + A_0 ~ \exp \big ( -B  ( \frac{ {\phi}} {\phi_{\text{3dB}}} )^2 \big )$, where $B=\ln(2)$, $\phi_{3\text{dB}}$ is the half-power beam-width, {$A_1$} and $A_0$ are two constant parameters \cite{ICASSPpaper}. Let $d_{ss}$, $d_{ps}$, $d_{s_t p_t}$ and $d_{sp}$  be the distances between \SUTx ~and \SURx, \PUTx ~and \SURx,   \PUTx ~and \SUTx, and \PURx ~and \SUTx,  respectively.
%
\par The fading coefficients from \SUTx ~to \SURx,  \PUTx ~to \SURx, \PUTx ~to \SUTx, and  \SUTx ~to \PURx ~are denoted by $g_{ss}$, $g_{ps}$, $g_{s_t p_t}$ and $g_{sp}$, respectively. We assume $g_{ss}$, $g_{ps}$, $g_{s_t p_t}$ and $g_{sp}$  are independent exponential random variables  with means $\gamma_{ss}$, $\gamma_{ps}$, $\gamma_{s_t p_t}$ and $\gamma_{sp}$, respectively. The path-loss is $L =  ({d_0}/{d})^\nu$, where $d_0$ is the reference distance, $d$ is the distance between users, and $\nu$  is the path loss exponent. We assume there is no cooperation between SUs and PUs and hence, \SUTx ~and \SURx ~only  do not know the realizations of  $g_{sp}$ and $g_{ps}$ and only know their statistics. On the other hand, \SUTx ~knows $g_{ss}$. 
\subsection{Spectrum Sensing}
\par The \SUTx ~employs a frame with duration $T$ seconds. Each frame consists of a sensing time slot with duration $\tau$ seconds (to be optimized) and \SUTx ~uses this time  to decide whether spectrum is idle or busy. The remaining frame of duration $T-\tau$ seconds is used for data transmission if the spectrum is sensed idle. It is clear that for a given $T$, if we increase the sensing time $\tau$, the spectrum  sensing will be more accurate. On the other hand, the available time for data transmission decreases. Therefore, a trade-off exists between the sensing time and the transmission capacity of our CR network.
\par We formulate the spectrum sensing at the \SUTx ~as a binary hypothesis testing problem where the received signal in \SUTx ~can be written as
\begin{equation}\label{E43}
\begin{array}{lll}
\mathcal{H}_0  & : & r[k] = w[k],\\
\mathcal{H}_1 & : & r[k] = \sqrt{g_{s_t p_t} A(\phi_t \!-\!\theta_{p_{t}}) L_{s_t p_t}} ~p[k] + w[k]\\
\end{array}%
\end{equation}
for $k=1, ..., N_s$. The two hypotheses  $\mathcal{H}_0$ and $\mathcal{H}_1$ with probabilities $\pi_0$ and $\pi_1 \! = \! 1 -\pi_0$ denote the spectrum is truly idle and truly busy, respectively. The term $w[k] \sim  \mathcal{N}(0, \sigma_n^2)$ is the additive white Gaussian noise (AWGN) at the \SUTx ~and $p[k]$ is  the transmitted symbol from the \PUTx ~with average power $P_p$. We assume the \SUTx ~knows $P_p$. We note $N_s=\tau f_s$ is the number of signal samples available for spectrum sensing and $f_s$ is the sampling frequency. Let $\widehat{\mathcal{H}}_1$ and $\widehat{\mathcal{H}}_0$ with probabilities $\pioh$ and $\pilh$ denote that the result of spectrum sensing is busy and idle, respectively. Considering energy detection as our spectrum sensing method, the decision statistics at the \SUTx ~can be written as $Z = \frac{1}{N_s}\sum_{k=1}^{N_s} \big | r[k] \big |^2$.
%
%
\par The accuracy of our spectrum sensing method is characterized  by false alarm probability  $P_f\!=\!\text{Pr}\{\widehat {\mathcal {H}}_1 | \mathcal {H}_0 \}$ and detection probability $P_d\!=\!\text{Pr}\{\widehat{\mathcal {H}}_1 | \mathcal{H}_1\}$.  For large $N_s$, we can use the central limit theorem and approximate the probability distribution function (PDF) of decision statistics $Z$ as Gaussian distribution and $P_f$ and $P_d$ can be written as \cite{R003}
\vspace{-1mm}
%
\begin{align}
P_f (\phi_t, \tau) = & ~Q\left( \Big(\frac{\xi}{\sigma^2_n} \!-\!1  \Big) \sqrt{\tau f_s}  \right) \label{Pf} \\
P_d (\phi_t, \tau)  = & ~Q\left( \Big(\frac{\xi}{\sigma^2_n}\!-\!\gamma\!-\!1  \Big) \sqrt{ \frac{\tau f_s}{2\gamma+1}}  \right) \label{Pd}
\end{align}
%
where $\gamma = P_p \gamma_{s_t p_t} A(\phi_t-\theta_{p_{t}}) L_{s_t p_t} /\sigma^2_n$ is the  signal-to-noise-ration (SNR) at the \SUTx ~and $\xi $ is the decision threshold. The probabilities  in \eqref{Pf} and \eqref{Pd} are functions of the optimization parameters $\tau$ and $\phi_t$. For the simplicity of the presentation, we drop the parameters $\tau$ and $\phi_t$ in the remaining of the paper.
\par The orientation of \SUTx's antenna ($\phi_t$) with respect to direction of \PUTx ~affects the spectrum sensing accuracy. To increase $P_d$ during spectrum sensing, the \SUTx's antenna should be pointed to \PUTx's direction to receive the maximum power. 
On the other hand, the \SUTx's antenna should be pointed to \SURx's direction to maximize the transmission capacity. Thus, there is  a sensing-capacity trade-off in terms of the \SUTx's antenna orientation. 
%
%
%
%
\vspace{-0mm}
\subsection{Data Communication Channel}
When the spectrum is sensed idle, the \SUTx ~uses power $P$ (to be optimized) to transmit signal to \SURx. Let $s[m]$ denote the transmitted signal by \SUTx ~with power $P$, and $y[m]$ denote the corresponding received signal by \SURx ~given by
\begin{equation*}\label{r_m}
y[m] = \sqrt{g_{ss} L_{ss} G(\theta,\phi_t,\phi_r)} ~s[m] + n[m], 
\end{equation*}
where $n[m]$ is the AWGN with power $\sigma_n^2$ and $G(\theta,\phi_t,\phi_r)=A(\phi_t\!-\!\theta) A(\phi_r\!-\!\pi\!-\!\theta)$ is the product of SU$_\text{tx}$  and SU$_\text{rx}$  antennas' gain. For the simplicity of presentation, we drop the parameters $\theta$, $\phi_t$ and $\phi_r$ from $G(\theta,\phi_t,\phi_r)$. 
\par Our goal is to find the ergodic capacity of the channel between \SUTx ~and \SURx  ~and explore the optimal \SUTx ~transmit power $P$, optimal sensing time $\tau$ and the optimal directions of \SUTx ~and \SURx ~antennas, $\phi_t$ and $\phi_r$, such that this capacity maximized, subject to peak transmit power and outage interference probability  constraints.
%
%
%
%
%
\vspace{-0mm}
\section{Constrained Ergodic Capacity  Maximization}\label{Se3}
\subsection{Capacity  Expression}
%
%
\par When spectrum sensing is imperfect, the ergodic capacity would depend  on the true status of the PU and the spectrum sensing result. In our problem, the ergodic capacity  becomes  $C= D  ~\mathbb{E} \Big \{ \alpha_{0} ~c_{0,0} + \beta_{0} ~c_{1,0}  \Big \}$, where $\mathbb{E}\{\cdot\}$ is the expectation operator, and $c_{i,0}$ is instantaneous capacity  corresponding to $\mathcal{H}_i$ and $\widehat{\mathcal{H}}_0$ with probability $\alpha_0 = \text{Pr}\{\mathcal{H}_0  ,\widehat{\mathcal{H}}_0\}$ and  $\beta_0 = \text{Pr}\{\mathcal{H}_1 ,\widehat{\mathcal{H}}_0\}$, given as
\vspace{-1.5mm}
%
\begin{align}\label{c0i}
c_{0,0} =  & ~\log_2 \left(1+\frac{g_{ss} L_{ss} G P}{\sigma^2_n}\right) \\
c_{1,0} =  & ~\log_2 \left(1+\frac{g_{ss} L_{ss} G P}{\sigma^2_n+P_p ~ g_{ps} L_{ps} ~A(\phi_r-\theta'_{p_t})}\right)
\end{align}
%
and $D \!=\! (T-\tau)/T$
is the fraction of time in which \SUTx ~transmits data to \SURx.
%
%
It is easy to verify $\alpha_0 = \pi_0  (1\!-\!P_f)$ and $\beta_0  = \pi_1  (1-P_d)$. It is worth noting that, if spectrum sensing errors are not considered (i.e., spectrum sensing is assumed to be perfect) $\alpha_0=\pi_0, \beta_0=0$. Also, it is important to emphasize that the optimal transmit power $P$, the optimal antenna directions and the optimal sensing time $\tau$ are functions of the fading coefficient  $g_{ss}$. 
After taking expectation with respect to $g_{sp}$ and $g_{ps}$, $C$ can be written as 
\vspace{-1mm}
%
\begin{equation}\label{Capacity_SU}
 C  \! = \! D ~\mathbb{E}_{g_{ss}} \!\! \left \{ \! \widehat{\pi}_0 \log_2 \! \big (1\!+\!\frac{1}{x} \big ) \! + \! \frac{\beta_0}{\ln(2)} \Big[  T({y}) \! - \! T \big({y}\! + \! \frac{{y}}{x} \big) \Big] \!  \right \}
\end{equation}
%
\noindent where $T(z)  =  e^{z} \mathrm{ Ei}\left( -z\right)$ and $\mathrm{Ei}(z)$  is the exponential integration \cite{BookTable}. In \eqref{Capacity_SU}, $x = {\sigma^2_n}/{a P}$, $a=g_{ss} L_{ss} G$,  ${y} = {\sigma^2_n}/{\bar{\sigma}^2_p}$. The term  $\bar{ \sigma}^2_p  = P_p \gamma_{ps} L_{ps} A(\phi_r-\theta_{p_t}')$ captures the interference on SU$_{\text{rx}}$ due to PU activities. 
%
%
%
%
%
\subsection{Constraints}
\vspace{-0.0mm}
Upon transmitting data, the \SUTx ~generates an interference on the \PURx. 
Similar to the outage concept developed in wireless communication community, we define the interference outage probability as the probability that the interference exceeds a maximum threshold $I_\text{pk}$. As a mechanism to control the interference generated by the \SUTx, we require that the interference outage probability to be smaller than a maximum value $\varepsilon$. In other words, we consider the following constraint
\vspace{-1mm}
%
\begin{equation}\label{Ip0}
\text{Pr} \Big \{  D \beta_0  P g_{s p} L_{sp} ~ A(\phi_t-\theta_{p_r}) > \Ipk \big| ~g_{ss} \Big \} \leq \varepsilon.
\end{equation}
%
%
We can rewrite \eqref{Ip0} as $F_{g_{sp}} \big (\frac{I_\text{pk}}{D   \beta_0 L_{sp}  A(\phi_t-\theta_{p_r})  P} \big) \geq 1-\varepsilon$, 
%
%
where $F_{g_{sp}} (\cdot)$  is the cumulative distribution function (CDF) of random variable $g_{sp}$, given as $ F_{g_{sp}}  (x) =  1-\exp({ \frac{-x}{\gamma_{sp} } }) $ for $x \geq 0$. Finally, we can write the constraint in \eqref{Ip0} as
\vspace{-1mm}
%
\begin{equation}\label{Iav}
D  \bar{b}_0 P \leq \frac{-\Ipk}{\ln(\varepsilon)}
\vspace{-1mm}
\end{equation}
%
\noindent where $\bar{b}_0 = \beta_0 \gamma_{sp} L_{sp} A(\phi_t-\theta_{p_r})$.  Let $P_\text{pk}$ indicate the maximum allowed instantaneous transmit power of  \SUTx.  To satisfy the peak transmit power constraint, we have
\vspace{-1mm}
\begin{equation}\label{Pav}
D \widehat{\pi}_0 P  \leq \Ppk.
\end{equation}
%
In order to ensure that the \SURx's  (\SUTx's) orientation lies within the  half-power beam-width of the \SUTx ~(\SURx) antenna, we constrain them as
\vspace{-1mm}
%
\begin{subequations}\label{angle_constraint}
\allowdisplaybreaks
\begin{align}
\big | \phi_t-\theta \big | \leq {\phi_{3\text{dB}}}, \label{phit_bound}\\
\big | \phi_r-\pi -\theta \big | \leq {\phi_{3\text{dB}}} \label{phir_bound}.
\end{align}
\vspace{-2mm}
\end{subequations}
\vspace{-4.0mm}
%
%
%
\subsection{Behavior of Capacity with respect to sensing time $\tau$}
\par In this section, we examine the behavior of $C$ with respect to $\tau$ and we show that, in a certain condition, $C$ is a concave function of $\tau$. Consider $P_f$, $P_d$ in \eqref{Pf}, \eqref{Pd}. By taking the second derivative of $P_f$ with respect to $\tau$, one can easily verify that $P_f$ is a convex function of $\tau$ when $\xi>\sigma^2_n$.  With the same argument we can show that $D \alpha_0$ is a concave functions with respect to $\tau$ for $\xi>\sigma^2_n$. Taking the first derivative of $C$ with respect to  $\tau$, we get
\vspace{-0mm}
%
%
\begin{equation*}
\frac{\partial C}{\partial \tau} =\frac{-C}{T\! -\! \tau}  + \frac{D}{\sqrt{8\pi \tau}} \mathbb{E}_{g_{ss}} \!  \bigg \{ \! X \pi_0 c_{0,0} ~e^{\frac{-\tau}{2} X^2  } \! + Y \pi_1 c_{1,0} ~e^{\frac{-\tau}{2} Y^2 } \! \bigg \}
\end{equation*}
%
%
%
where $X =  \big (\frac{\xi}{\sigma^2_n} \! -\!1 \big ) \sqrt{ f_s}$ and $ Y =  \big (\frac{\xi}{\sigma^2_n} \! -\! \gamma\!-\!1  \big ) \sqrt{\frac{ f_s}{2\gamma+1}}$.
%
%
%
\noindent One can easily verify that  $\underset{\tau \to T} \lim ~{\partial C}/{\partial \tau} < 0.$
If we choose $ \xi \geq \sigma^2_n(1+ m \gamma)$ where $m = \frac{\pi_1}{\pi_1 + \pi_0 \sqrt{2 \gamma + 1}} < 1$,
we get $\underset{\tau \to 0} \lim ~{\partial C}/{\partial \tau} \to +\infty$.
We conclude that $C$ increases when $\tau$ goes to zero and decreases when $\tau$ goes to $T$. Hence, $C$ has a maximum point with respect to $\tau$ within the interval $(0, T)$. However, $C$ may not be concave with respect to $\tau$ and we may need to a use a numerical search method to find the optimal $\tau$. 
On the other hand, it is easy to verify that $P_d$ is a concave function of $\tau$ when $\xi<\sigma^2_n(1+ \gamma)$. We conclude that if we choose the decision threshold $\xi$ such that $\sigma^2_n(1+ m \gamma) \leq \xi < \sigma^2_n(1+ \gamma)$, the capacity $C$ will be a concave function of $\tau$. 
%
%
%
\subsection{Solution}
\vspace{-0mm}
\par Recall that our goal is to maximize  the ergodic capacity $C$ over $P$, $\phi_t$, $\phi_r$ and $\tau$ subject to constraints \eqref{Iav}, \eqref{Pav} and \eqref{angle_constraint}. The capacity is concave with respect to $P$ and $\phi_r$. However, in general, it is not concave with respect to $\phi_t$ and $\tau$. 
In the following we present our approach for solving this optimization problem. The optimal power can be written as 
\vspace{-1mm}
\begin{equation}\label{Popt}
P^{\text{opt}} = \text{min} ~\Big \{ \frac{P_\text{pk}}{D \widehat{\pi}_0} ~ , ~ \frac{-\Ipk}{D \bar{b}_0 \ln(\varepsilon)} \Big \}.
\end{equation}
The optimal $\phi_t$ and $\tau$ can be obtained by using searching methods like bisection method. We consider an initial value for $\phi_t$ which satisfies \eqref{phit_bound} and $\tau \in (0,T)$, and obtain $P^{\text{opt}}$ using \eqref{Popt}. Then, the optimal $\phi_r$ can be found by solving ${\partial C}/{\partial \phi_r}=0$, subject to the constraint \eqref{phir_bound}. For any realization of $g_{ss}$, the first derivative of {$C$} with respect to $\phi_r$ is equal to
\vspace{-1mm}
%
\begin{equation*}
\frac{D}{\ln(2)} \! \left \{ \! \frac{A'(\phi_r \! -\!\pi\! - \!\theta)}{A(\phi_r \! -\!\pi\! - \!\theta)} g_0(x,y) \! - \! \frac{A'(\phi_r \! - \!\theta'_{p_t})}{A(\phi_r \! - \!\theta'_{p_t})} k_0(x,y) \right \}
\end{equation*}
%
\vspace{-1mm}
where
\vspace{-1mm}
%
\begin{align*}
 g_0(x,y) = & \frac{\pioh+2 \beta_0 }{2(1+x)} + \frac{{y}\beta_0}{x}   T \big ({y}+\frac{{y}}{x} \big ), \\
 k_0(x,y) = & \beta_0 \Big [ 2+ {y} T({y}) + {y}(1+\frac{1}{x})   T \big ({y}+\frac{{y}}{x} \big ) \Big ],
\end{align*}
%
and $A'(\cdot) = {\partial A(\cdot)}/{\partial \phi_r}$. Then, we find the value of $\phi_t$ and $\tau$ which maximizes $C$. Algorithm \ref{Alg} summarizes our proposed approach to find the optimal solutions $\tau^\text{opt}$, $\phi_t^ \text{opt}$, $\phi_r^ \text{opt}$ and $P^\text{opt}$.
%
%
\begin{algorithm}[t]
\caption{{Optimization Algorithm}}
\label{Alg}
$\phi_t^{(0)}=\phi_{\text{init}}$ which satisfies \eqref{phit_bound}\\
$\tau^{(0)}=\tau_{\text{init}} \in (0, T)$\\
calculate $P$ using \eqref{Popt}.\\
solve $\partial C/\partial \phi_r = 0$ and obtain $\phi_r$.\\
$[\phi_t^{\text{opt}} , \tau^{\text{opt}} ]= \text{argmax} \left\{C \right\}$ using bisection search\\

$P^\text{opt} = [P]_{\phi_t=\phi_t^{\text{opt}},  ~{\tau=\tau^{\text{opt}}} }$\\
$\phi_r^\text{opt} = [\phi_r]_{\phi_t=\phi_t^{\text{opt}}, ~{\tau=\tau^{\text{opt}}} }$\\
\vspace{-0mm}
\end{algorithm}
%
\vspace{-0mm}
\section{Numerical Results and Conclusion}\label{}
\begin{figure}[!t]
\vspace{-0mm}
\centering
	\begin{subfigure}[b]{0.25\textwidth}                
		\psfrag{tau [ms]}[Bl][Bl][0.6]{$\tau$ [ms]}
		\psfrag{Pp=0.1 --------}[Bl][Bl][0.45]{$P_p=0.1$ W}
		\psfrag{Pp=05 --------}[Bl][Bl][0.45]{$P_p=~5$ W}
		\psfrag{Pp=15 --------}[Bl][Bl][0.45]{$P_p=15$ W}
		\psfrag{Ergodic Capacity}[Bl][Bl][0.6]{~~~~Capacity}
		\includegraphics[width=42mm]{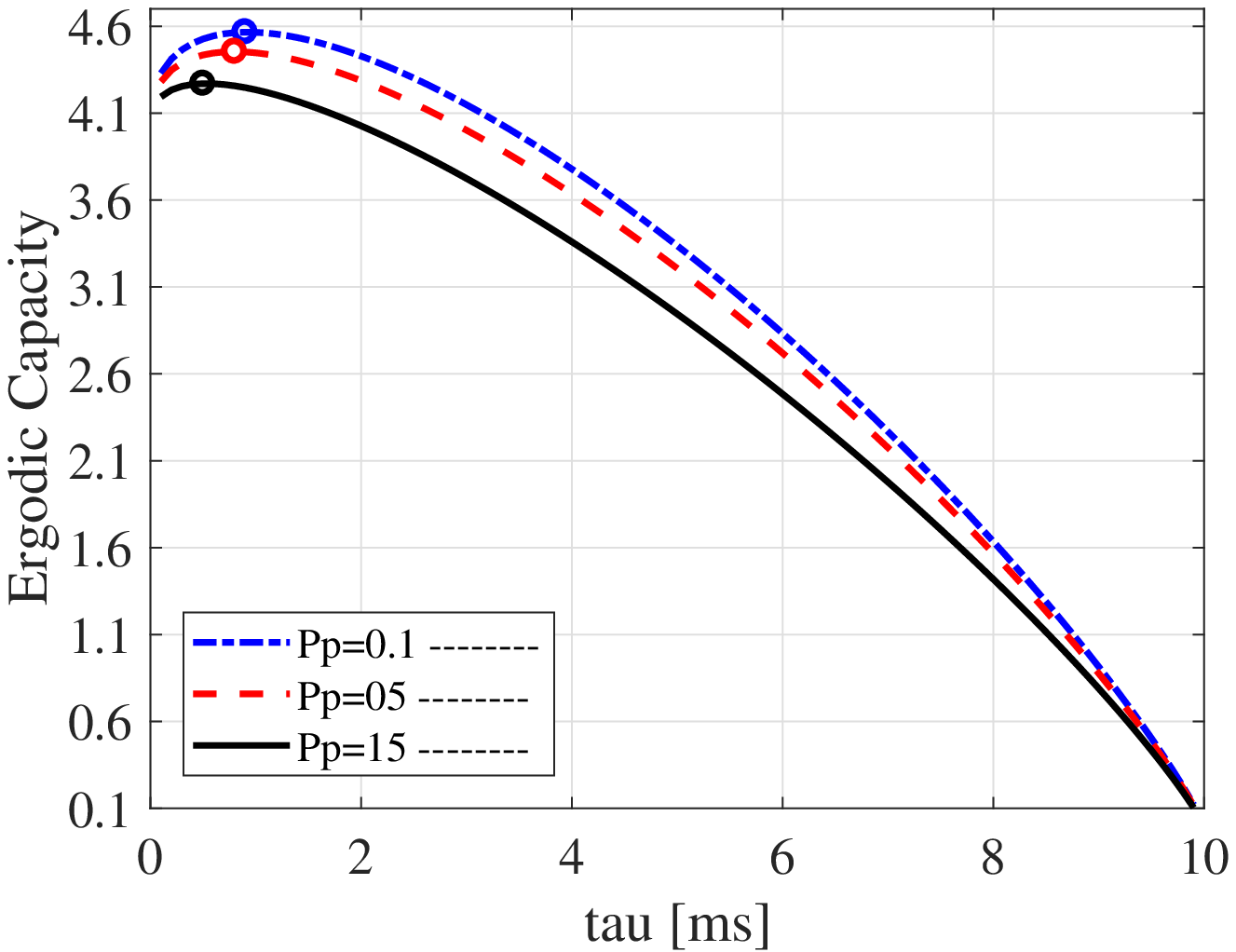}
		\caption{Variations of $C$ versus $\tau$.} 
		\label{Capacity_tau_fig}    	         
	\end{subfigure}%
      \begin{subfigure}[b]{0.25\textwidth}
		\psfrag{epsilon}[Bl][Bl][0.6]{$\varepsilon$}
		\psfrag{Ergodic Capacity}[Bl][Bl][0.6]{~~~~~$C_\text{opt}^\text{Dir}$}
		\includegraphics[width=42.5mm]{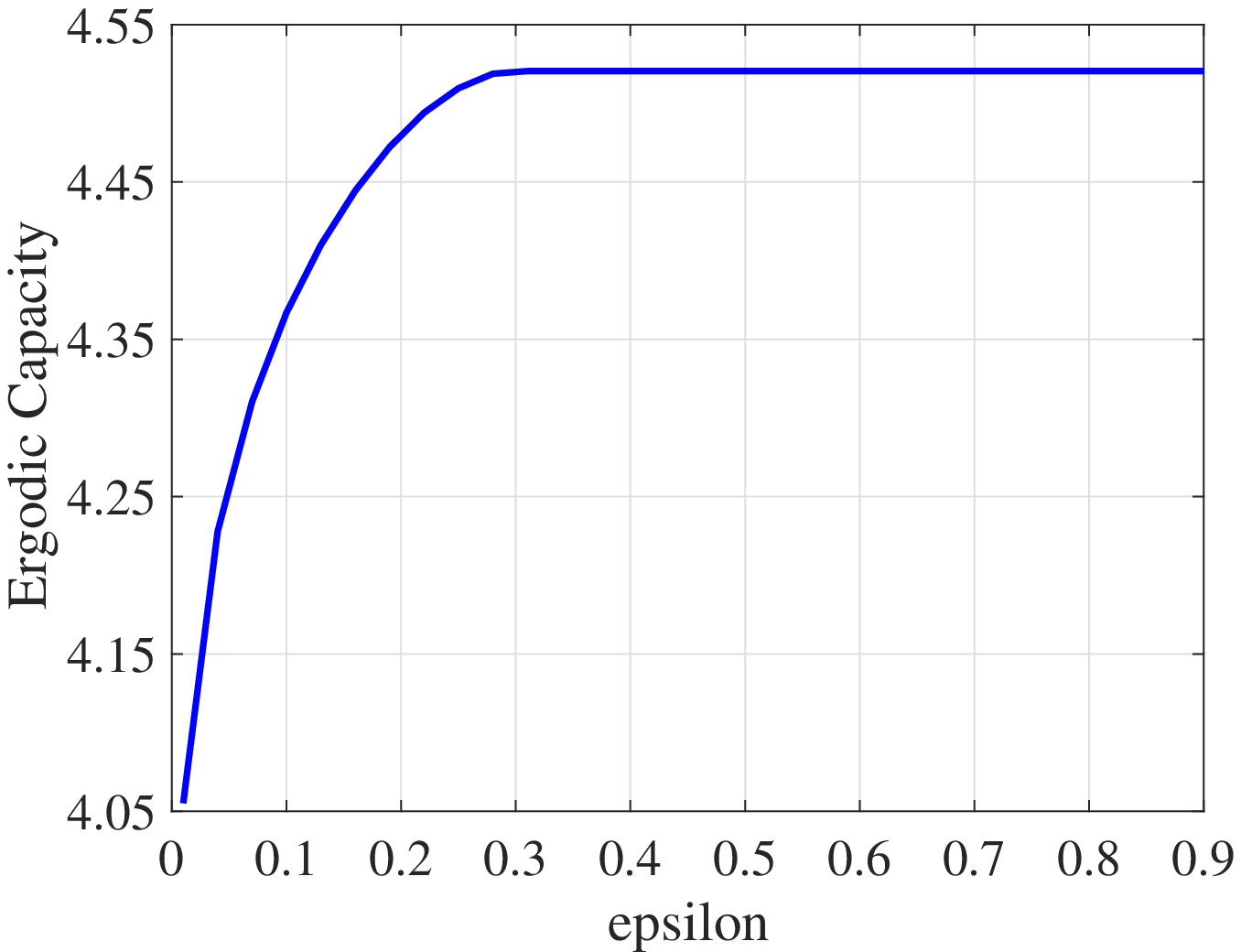}	
		\caption{Variations of $C_\text{opt}^\text{Dir}$ versus $\varepsilon$.} 
		\label{Capacity_epsilon}    
      \end{subfigure} \\
     \vspace{4mm}
     \begin{subfigure}[b]{0.25\textwidth}
		\centering
		\psfrag{theta}[Bl][Bl][0.6]{$\theta$ [degree]}
		\psfrag{eps=0.05}[Bl][Bl][0.45]{$\varepsilon=0.05$}
		\psfrag{eps=0.1}[Bl][Bl][0.45]{$\varepsilon=0.1$}
		\psfrag{eps=0.2}[Bl][Bl][0.45]{$\varepsilon=0.2$}
		\psfrag{Ergodic Capacity}[Bl][Bl][0.6]{~~~~~$C_\text{opt}^\text{Dir}$}
		\includegraphics[width=42mm]{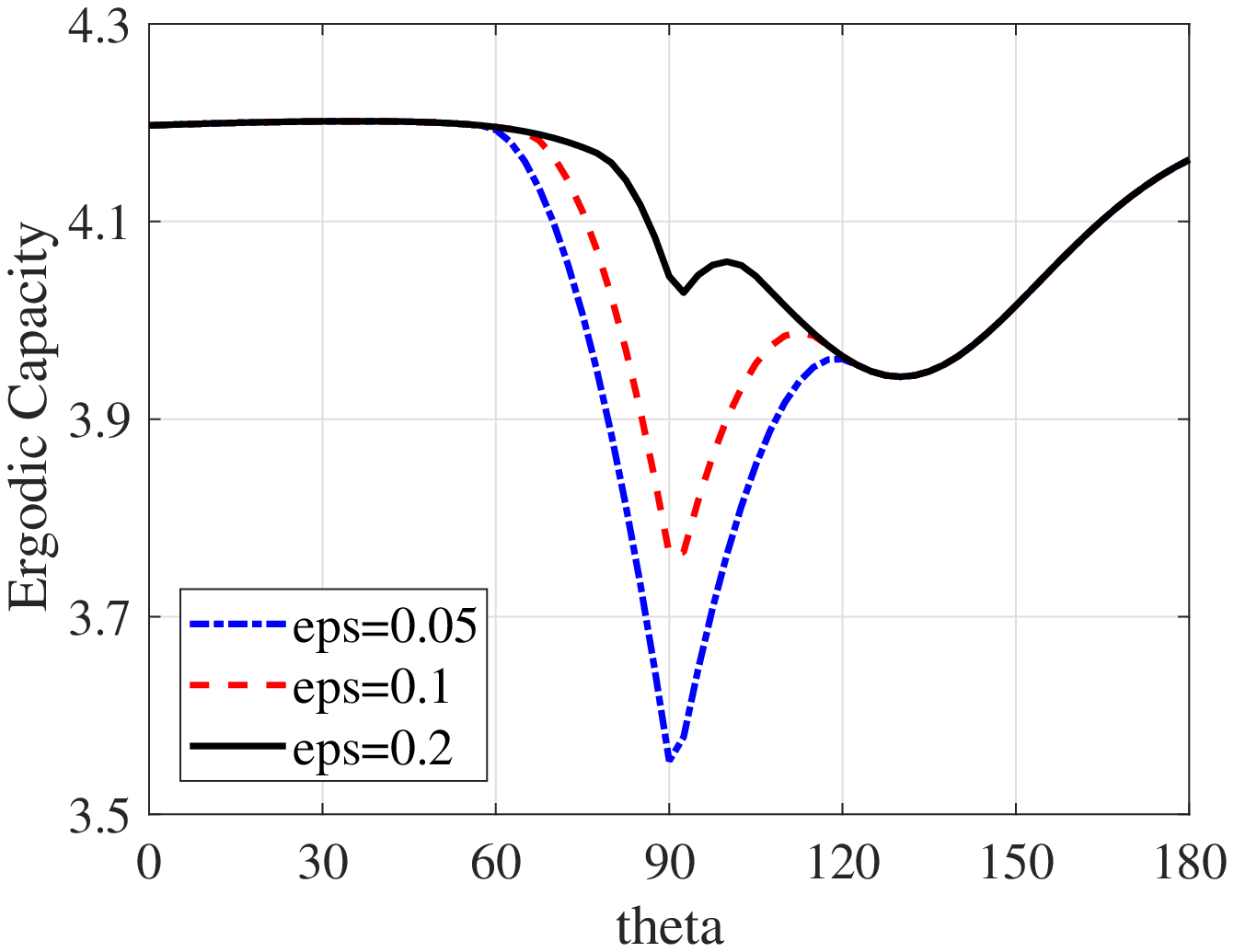}
		\caption{$C_\text{opt}^\text{Dir}$ versus $\theta$  for $\varepsilon=0.05, 0.1, 0.2$.} 
		\label{Capacity_theta_epsilon}   
     \end{subfigure}%
\caption{ Variations of capacity versus (a) $\tau$, (b) $\varepsilon$ and (c) $\theta$.}
\vspace{-0mm}
\end{figure}

In this section, we illustrate how effectively the directional antennas can improve the capacity of our secondary network  by Matlab simulations. Assume $\sigma^2_n\!=\!1$, $A_0\! =\! 9.8$, {$A_1\!=\!0.2$}, $\gamma_{ss}\!= \!\gamma_{sp}\!= \!\gamma_{ps}\!=\!\gamma_{p_t s_t}\! = \!1$, $\pi_1\! =\! 0.3$, $T=10$ ms, $f_s = 20$ KHz, $\phi_{3\text{dB}}\!=\!30 \degree$, $\varepsilon = 0.05$, $\theta_{p_r}\!=\!90\degree$, $P_p \!=\!0.4$ watts and ${\theta^\prime}_{p_t}\!=\!130\degree$. Fig. \ref{Capacity_tau_fig} depicts $C$ versus $\tau$ for different values of $P_p$ when $\Ipk \!= \!2$ dB, $\Ppk \!= \!10$ dB and $\theta \! =\! 50\degree$. We can see that $C$ always has a maximum in the interval $(0, T)$. Also, we can see that as the power of \PUTx ~($P_p$) increases, $\tau^{\text{opt}}$ decreases and \SUTx ~needs less time to sense the activity of the \PUTx.
%
%
We denote  the optimal  capacity of CR network using directional antennas by $C_\text{opt}^\text{Dir}$ .  Fig. \ref{Capacity_epsilon} plots $C_\text{ opt}^\text{Dir}$  versus $\varepsilon$, when $P_\text{pk} \!= \!10$ dB and $\Ipk \!= \!2$ dB.  By increasing $\varepsilon$, the \SUTx ~can use  a higher power level to transmit data to the \SURx ~without violating the interference outage probability constraint in \eqref{Iav}. As a result, $C_\text{opt}^\text{Dir}$ increases. We observe  that when  $\varepsilon<0.28$, the interference outage probability constraint in \eqref{Iav} is dominant. However, for  $\varepsilon \geq 0.28$, the instantaneous power constraint in \eqref{Pav} is dominant and $C_\text{opt}^\text{Dir}$ will not increase by increasing $\varepsilon$.
%
%
%
%
%
%
\begin{figure}[ht] 
\vspace{-0mm}
  \begin{subfigure}[b]{0.5\linewidth}
		\centering
		\psfrag{theta}[Bl][Bl][0.6]{$\theta$ [degree]}
		\psfrag{Pp = 0.4  W}[Bl][Bl][0.45]{$P_p=0.4$ W}
		\psfrag{Pp = 1.2  W}[Bl][Bl][0.45]{$P_p=1.2$ W}
		\psfrag{Ergodic Capacity}[Bl][Bl][0.6]{~~~~~$C_\text{opt}^\text{Dir}$}
		\includegraphics[width=42mm]{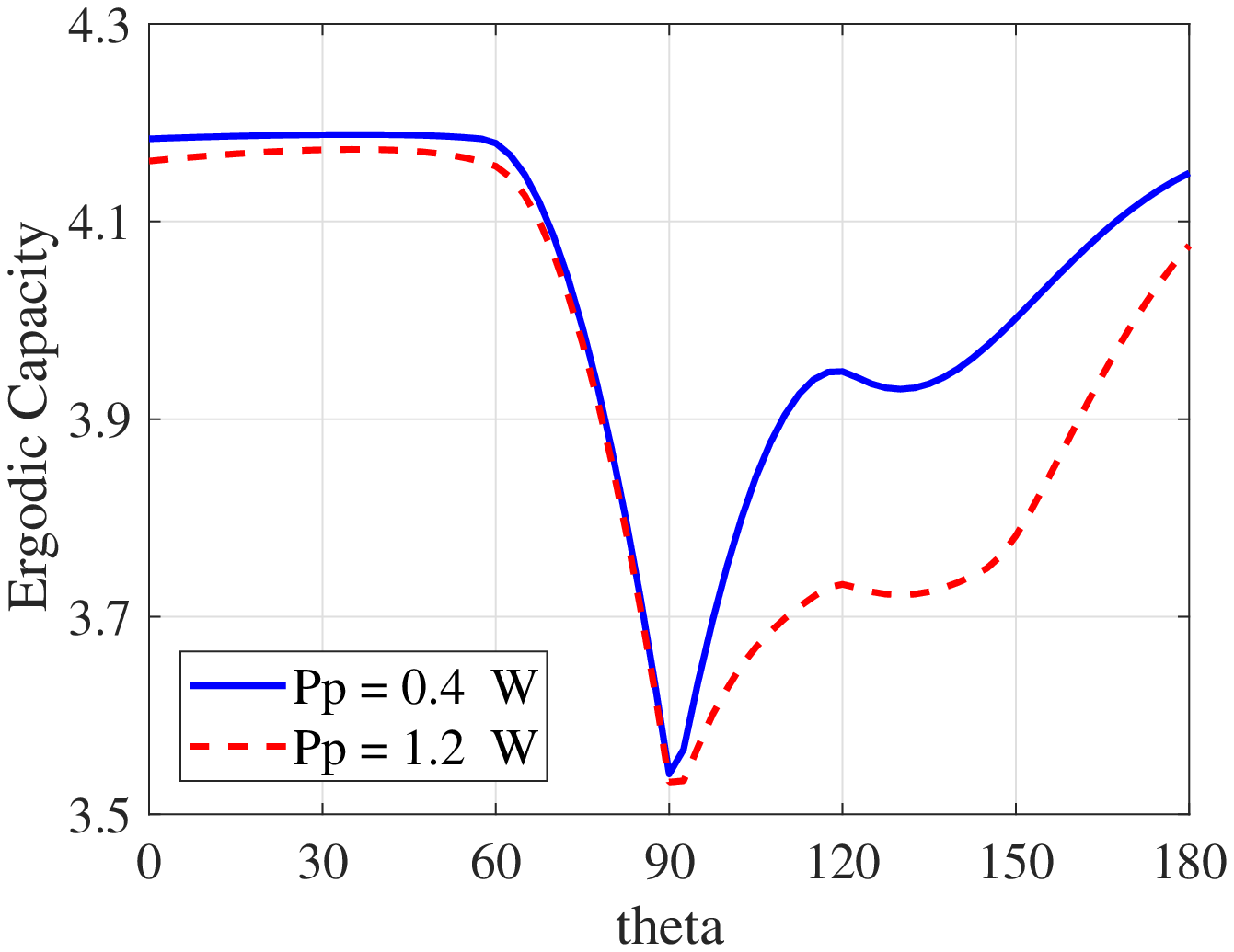}
		\caption{$C_\text{opt}^\text{Dir}$ versus $\theta$ for $P_p=0.4, 1.2$ W.} 
		\label{Capacity_theta_Pp}  
	      \vspace{5ex}
  \end{subfigure}
  \begin{subfigure}[b]{0.5\linewidth}
		\centering
		\psfrag{theta}[Bl][Bl][0.6]{$\theta$ [degree]}
		\psfrag{phi3dB = 30}[Bl][Bl][0.45]{$\phi_{3\text{dB}}=30\degree$}
		\psfrag{phi3dB = 45}[Bl][Bl][0.45]{$\phi_{3\text{dB}}=45\degree$}
		\psfrag{Ergodic Capacity}[Bl][Bl][0.6]{~~~~~$C_\text{opt}^\text{Dir}$}
		\includegraphics[width=42mm]{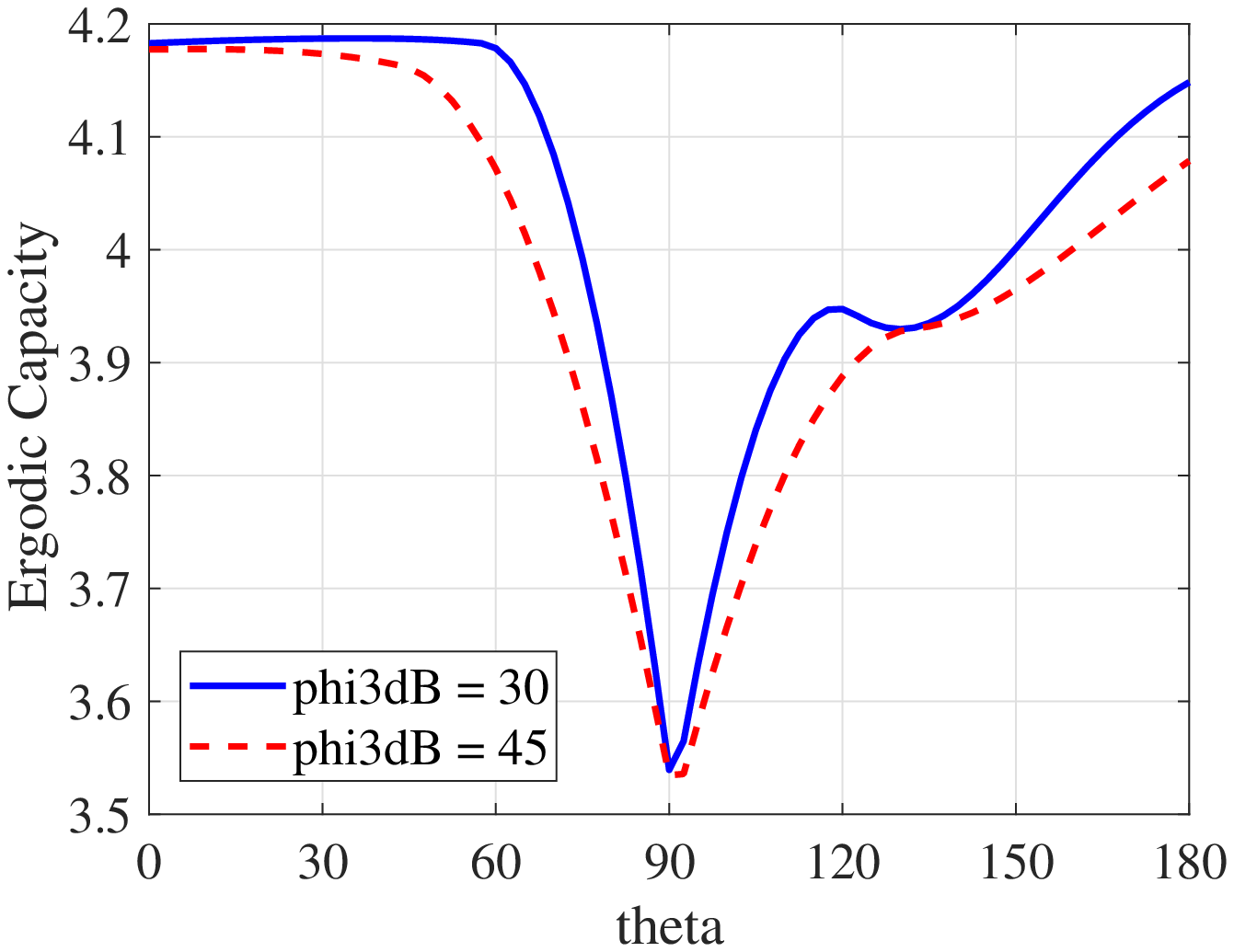}
		\caption{$C_\text{opt}^\text{Dir}$ versus $\theta$ for $\phi_{3\text{dB}}=30\degree, 45\degree$.} 
		\label{Capacity_theta_phi3dB}  
		\vspace{5ex} 
  \end{subfigure} 
  \begin{subfigure}[b]{0.5\linewidth}
		\vspace{-4mm}
		\centering
 		\psfrag{theta}[Bl][Bl][0.6]{$\theta$ [degree]}
		\psfrag{Dirc}[Bl][Bl][0.35]{$C_\text{opt}^\text{Dir}$}
		\psfrag{LOS}[Bl][Bl][0.35]{$C_\text{opt}^\text{LOS}$}
		\psfrag{Omni}[Bl][Bl][0.35]{$C_\text{opt}^\text{Omn}$}
		\psfrag{Ergodic Capacity}[Bl][Bl][0.6]{Ergodic Capacity}
		\includegraphics[width=42.0mm]{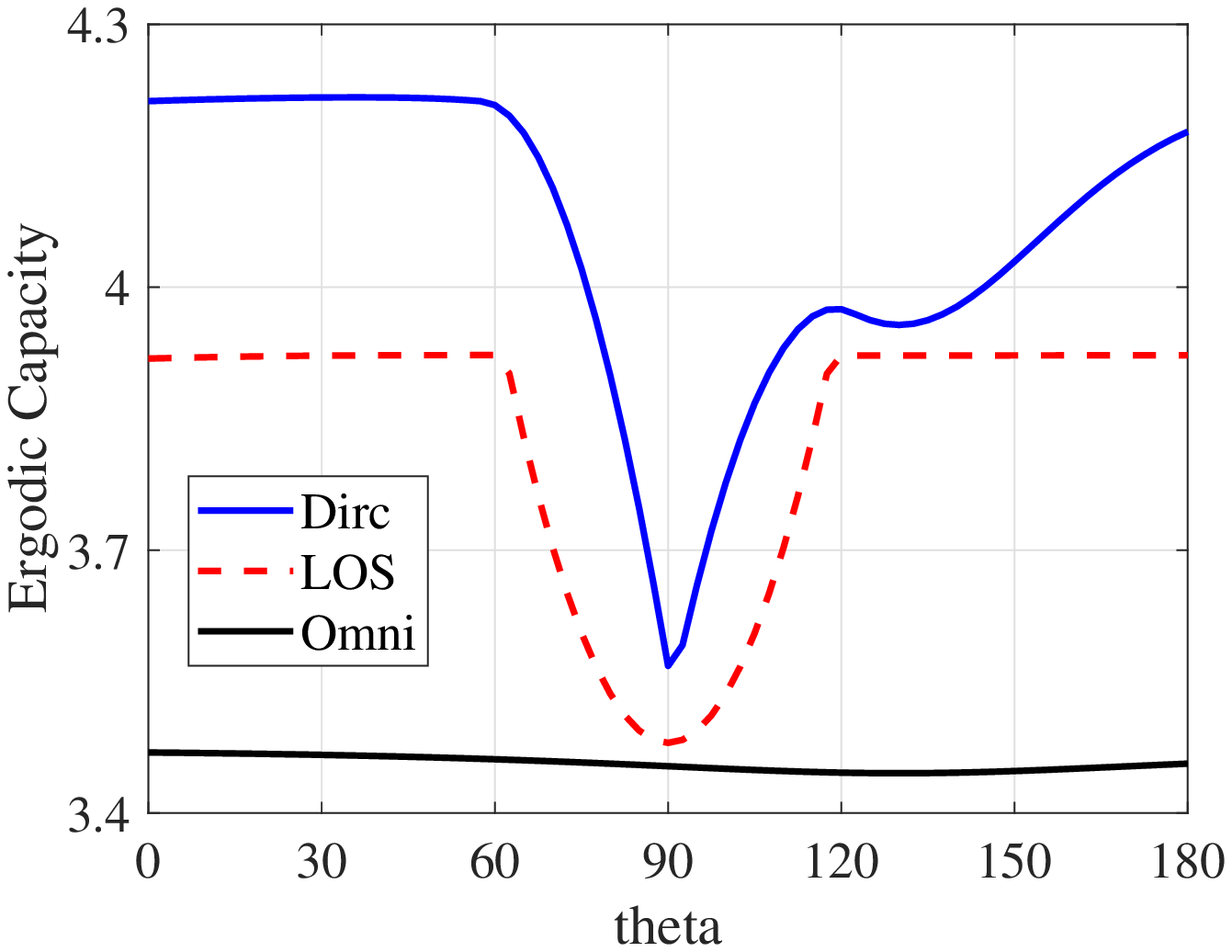}
		\caption{$C_\text{opt}^\text{Dir}$, $C_\text{opt}^\text{LOS}$ and  $C_\text{opt}^\text{Omn}$ versus $\theta$.} 
		\label{C_Dirc_LOS}    
  \vspace{+8mm}
  \end{subfigure}
  \begin{subfigure}[b]{0.5\linewidth}
		\centering
		\vspace{-4mm}
		\psfrag{theta}[Bl][Bl][0.6]{$\theta$ [degree]}
		\psfrag{Ppk = 6 dB}[Bl][Bl][0.4]{$P_\text{pk} = 6$ dB}
		\psfrag{Ppk = 8 dB}[Bl][Bl][0.4]{$P_\text{pk} = 8$ dB}
		\psfrag{Capacity Ratio}[Bl][Bl][0.6]{\!\!\!\!\!\!\!\! Capacity Ratio ~($\Gamma_\text{D2O}$)}
		\includegraphics[width=42mm]{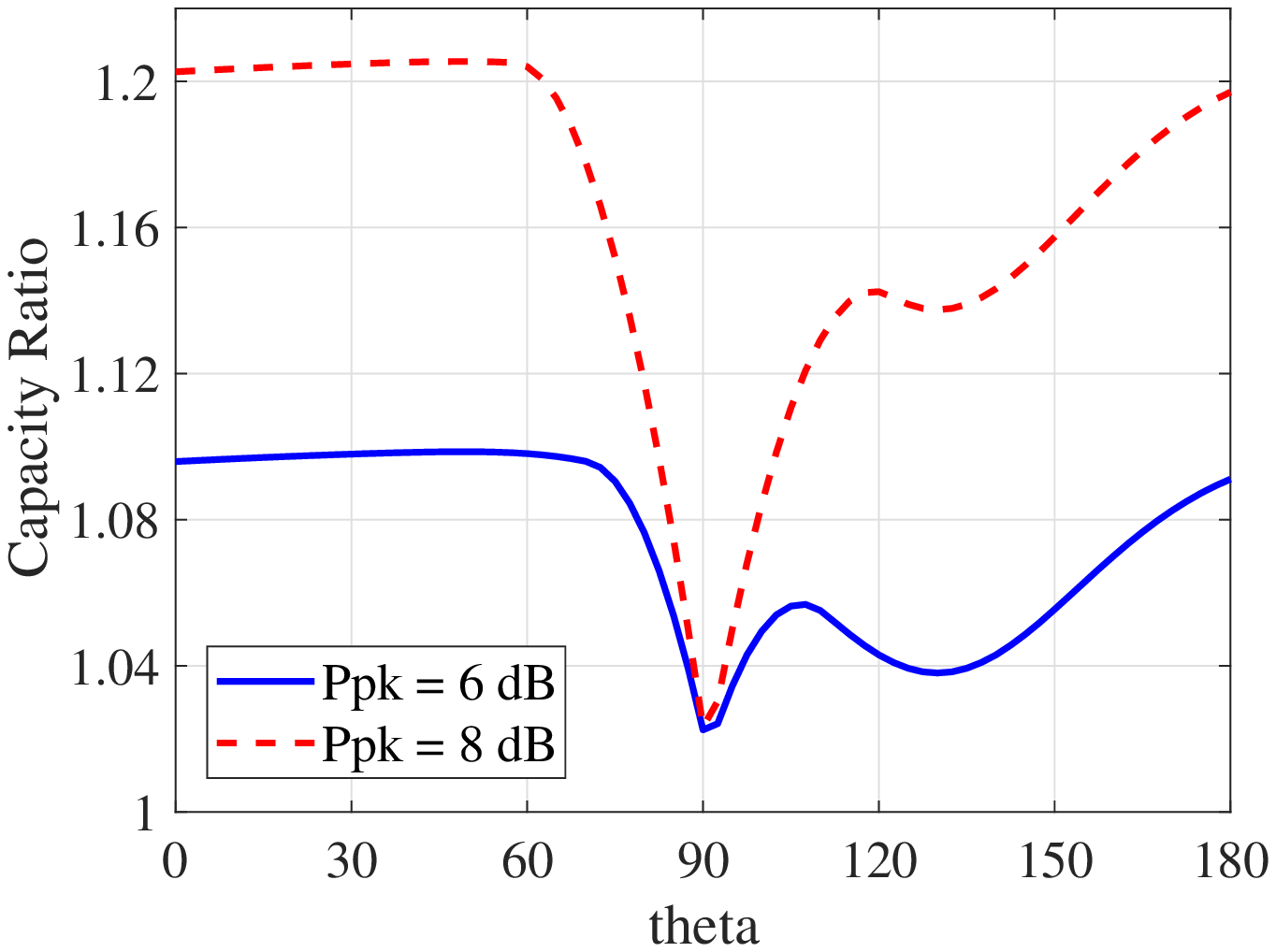}
		\caption{$\Gamma_\text{D2O}$ versus $\theta$ for $P_\text{pk} = 6, 8$ dB.} 
		\label{Capacity_Dirc2Omni_Ppk}    
  \vspace{+8.5mm}      
  \end{subfigure} 
  \vspace{-12mm}
  \caption{(a)--(c) Capacity $C_\text{opt}^\text{Dir}$ versus $\theta$ for different system parameters. (d) Capacity ratio $\Gamma_\text{D2O}$ versus $\theta$ for different $P_\text{pk}$.}
\vspace{-3mm}
\end{figure}
%
\par Fig. \ref{Capacity_theta_epsilon} illustrates $C_\text{opt}^\text{Dir}$ versus the orientation of the \SURx ~($\theta$) for $\varepsilon=0.05, 0.1, 0.2$ is illustrated. In this figure, we fix the orientation of \PUTx, \PURx ~and \SUTx ~and inspect the variations of $C_\text{opt}^\text{Dir}$  when the \SURx ~changes its location. For a specific value of $\varepsilon$, when the \SURx ~get closer to the \PURx, the \SUTx ~decreases $P$ and rotates  the main lobe of its antenna to satisfy the interference outage probability constraint. As a result, the $C_\text{opt}^\text{Dir}$ decreases.  When the \SURx ~gets farther away from \PURx ~($\phi_t \! > \! 90\degree$), $C_\text{opt}^\text{Dir}$ slightly increases. However, when $\phi_t \to \theta_{p_t}(=130\degree)$, $C_\text{opt}^\text{Dir}$  decreases again since the interference from the \PUTx ~to the \SURx ~($\bar{ \sigma}^2_p$) increases.
\par Fig. \ref{Capacity_theta_Pp} depicts $C_\text{opt}^\text{Dir}$ versus $\theta$ for two values of $P_p$. Increasing $P_p$, enhances the accuracy of spectrum sensing. On the other hand, it incurs stronger interference on the \SURx ~and reduces $C_\text{opt}^\text{Dir}$. Fig. \ref{Capacity_theta_phi3dB} shows the effect of the half-power beam-width ($\phi_{3\text{dB}}$) of directional antennas on $C_\text{opt}^\text{Dir}$. We can see that an antenna with narrower beam-width always yields higher capacity because it can cancel more interference from (to) \PUTx ~(\PURx).
\par Let $C_{\text{opt}} ^{\text{Omn}}$ denote the optimal  capacity when \SUTx ~and \SURx ~have omni-directional antennas  and only transmit power $P$ and sensing time  $\tau$ are optimized subject to constraints \eqref{Iav} and \eqref{Pav}.  Furthermore, let  $C_{\text{opt}} ^{\text {LOS}}$ be the optimal  capacity when directional antennas of \SUTx ~and \SURx ~are exactly pointed at each other and only $P$ and $\tau$ are optimized subject to constraints \eqref{Iav} and \eqref{Pav}. Fig. \ref{C_Dirc_LOS} illustrates $C_\text{opt}^\text{Dir}$, $C_\text{opt}^\text{LOS}$ and  $C_\text{opt}^\text{Omn}$ versus $\theta$. This figure shows the effectiveness of using the directional antennas and as well as optimizing their orientation on the capacity of the secondary network. It demonstrates that directional antennas can improve the secondary network capacity for all values of $\theta$.
\par We define the  capacity ratio $\Gamma_\text{D2O} = C_{\text{opt}}^{\text{Dir}} / C_{\text{opt}}^{\text{Omn}}$. This ratio is shown in Fig. \ref{Capacity_Dirc2Omni_Ppk} for $P_\text{pk} \! = \! 6, 8$ dB. We see that directional antennas yields  as much as 22\%  capacity gain for $P_\text{pk}\! =\! 8$ dB in comparison to omni-directional antennas. Also, we can see that the capacity gain decreases when maximum allowable transmit power ($P_\text{pk}$) decreases.
\par In summary,  we considered a CR system, where the SUs are equipped with directional antennas and sense the spectrum for duration of $\tau$. We formulated the ergodic capacity of the secondary network which uses energy detection method for spectrum sensing. The optimal \SUTx ~transmit power, the optimal sensing time $\tau$ and the optimal directions of \SUTx ~transmit antenna and \SURx ~receive antenna are obtained by maximizing the ergodic capacity, subject to peak transmit power and outage interference probability constraints. Our simulation results demonstrated the effectiveness of these optimizations on increasing the ergodic capacity of the secondary network.
%
%
%
%
%
\section*{Acknowledgment}
This research is supported by NSF under grant ECCS-1443942.
%
\bibliographystyle{IEEEtran}
\bibliography{CISSMyRef.bib}
%
\end{document}